\documentclass[aps,superscriptaddress,tightenlines,nofootinbib]{revtex4}
\usepackage{graphicx,amssymb,amsbsy,bm,amsmath}
\usepackage{hyperref}

\newcommand{\be}{\begin{equation}}
\newcommand{\ee}{\end{equation}}
\newcommand{\bea}{\begin{eqnarray}}
\newcommand{\eea}{\end{eqnarray}}
\begin{document}
\title{Quantum corrections to the inflaton potential and the power spectra
from superhorizon modes and trace anomalies.}
\author{D. Boyanovsky}
\email{boyan@pitt.edu} \affiliation{Department of Physics and
Astronomy, University of Pittsburgh, Pittsburgh, Pennsylvania 15260,
USA} \affiliation{Observatoire de Paris, LERMA. Laboratoire
Associ\'e au CNRS UMR 8112.
 \\61, Avenue de l'Observatoire, 75014 Paris, France.}
\affiliation{LPTHE, Universit\'e Pierre et Marie Curie (Paris VI) et
Denis Diderot (Paris VII), Laboratoire Associ\'e au CNRS UMR 7589,
Tour 24, 5\`eme. \'etage, 4, Place Jussieu, 75252 Paris, Cedex 05,
France}
\author{H. J. de Vega}
\email{devega@lpthe.jussieu.fr} \affiliation{LPTHE, Universit\'e
Pierre et Marie Curie (Paris VI) et Denis Diderot (Paris VII),
Laboratoire Associ\'e au CNRS UMR 7589, Tour 24, 5\`eme. \'etage, 4,
Place Jussieu, 75252 Paris, Cedex 05,
France}\affiliation{Observatoire de Paris, LERMA. Laboratoire
Associ\'e au CNRS UMR 8112.
 \\61, Avenue de l'Observatoire, 75014 Paris, France.}
\affiliation{Department of Physics and Astronomy, University of
Pittsburgh, Pittsburgh, Pennsylvania 15260, USA}
\author{N. G. Sanchez}
\email{Norma.Sanchez@obspm.fr} \affiliation{Observatoire de Paris,
LERMA. Laboratoire Associ\'e au CNRS UMR 8112.
 \\61, Avenue de l'Observatoire, 75014 Paris, France.}
\date{\today}
\begin{abstract}
We obtain the \emph{effective} inflaton potential during slow roll
inflation by including the one loop quantum corrections to the
energy momentum tensor from scalar curvature and tensor perturbations as
well as quantum fluctuations from light scalars and light Dirac
fermions generically coupled to the inflaton.
During slow roll inflation there is a clean and unambiguous separation
between superhorizon and subhorizon contributions to the energy
momentum tensor. The superhorizon part is determined by the
curvature perturbations and scalar field fluctuations: both
feature infrared enhancements as the inverse of a combination of slow roll
parameters which measure the departure from scale invariance in each case.
Fermions and gravitons do not exhibit infrared divergences.
The subhorizon part is completely specified by the trace anomaly
of the fields with different spins and is solely determined by the
space-time geometry. The one-loop quantum corrections to the amplitude of 
curvature and tensor perturbations are obtained to leading order in slow-roll 
and in the $ (H/M_{PL})^2$ expansion. This study provides a complete
assessment of the backreaction problem up to one loop 
including bosonic and fermionic degrees of freedom. The result
validates the effective field theory description of inflation and
confirms the robustness of the inflationary paradigm to quantum
fluctuations. Quantum corrections to the power spectra are 
expressed in terms of the CMB observables: $ n_s, \; r $ and $ dn_s/d \ln k $.
Trace anomalies (especially the graviton part) dominate these quantum
corrections in a definite direction: they {\bf enhance} the
scalar curvature fluctuations and {\bf reduce} the tensor fluctuations.

\end{abstract}


\maketitle
\tableofcontents
\section{Introduction}

Inflation is a central part of early Universe cosmology passing
many observational tests and becoming a predictive scenario
scrutinized by current and forthcoming observations. Inflation was
introduced to solve several shortcomings of the standard Big Bang
cosmology\cite{guth}-\cite{riottorev}. It provides a mechanism for
generating scalar (density) and tensor (gravitational wave)
perturbations\cite{mukha}-\cite{mukhanov}. A distinct aspect of
inflationary perturbations is that these are generated by quantum
fluctuations of the scalar field(s) that drive inflation. After
their wavelength becomes larger than the Hubble radius, these
fluctuations are amplified and grow, becoming classical and
decoupling from  causal microphysical processes. Upon re-entering
the horizon, during the matter era, these classical perturbations
seed the inhomogeneities which generate structure upon
gravitational collapse\cite{mukha}-\cite{mukhanov}. A great
diversity of inflationary models predict fairly generic features:
a gaussian, nearly scale invariant spectrum of (mostly) adiabatic
scalar and tensor primordial fluctuations, making the inflationary
paradigm fairly robust. The gaussian, adiabatic and nearly scale
invariant spectrum of primordial fluctuations provide an excellent
fit to the highly precise wealth of data provided by the Wilkinson
Microwave Anisotropy Probe
(WMAP)\cite{komatsu,spergel,kogut,peiris}. Perhaps the most
striking validation of inflation as a mechanism for generating
\emph{superhorizon} (`acausal')  fluctuations is the
anticorrelation peak in the temperature-polarization (TE) angular
power spectrum at $l \sim 150$ corresponding to superhorizon
scales\cite{kogut,peiris}.

The confirmation of many of the robust predictions of inflation by
current high precision observations places
inflationary cosmology on solid grounds. Forthcoming observations
will begin to discriminate among different inflationary models, placing
stringent constraints on them.
There are small but important telltale discriminants
amongst different models: non-gaussianity, a running spectral
index for scalar and tensor perturbations, an
isocurvature component for scalar perturbations, the ratios
for the amplitudes between tensor and scalar modes, etc. Already
WMAP reports a hint of deviations from constant scaling exponents
(running spectral index) and rules out the purely
monomial $\Phi^4$ potential\cite{peiris}.

Amongst the wide variety of inflationary scenarios, single field
\emph{slow roll} models\cite{barrow,stewlyth} provide an
appealing, simple and fairly generic description of inflation. Its
simplest implementation is based on a scalar field (the inflaton)
whose homogeneous expectation value drives the dynamics of the
scale factor, plus small quantum fluctuations. The inflaton
potential, is fairly flat during inflation. This flatness not only
leads to a slowly varying Hubble parameter, hence ensuring a
sufficient number of e-folds, but also provides an explanation for
the gaussianity of the fluctuations as well as for the (almost)
scale invariance of their power spectrum. A flat potential
precludes large non-linearities in the dynamics of the
\emph{fluctuations} of the scalar field. The current WMAP data
seems to validate the simpler one-field slow roll
scenario\cite{peiris}. Furthermore, because the potential is flat
the scalar field is almost massless, and modes cross the horizon
with an amplitude proportional to the Hubble parameter. This fact
combined with a slowly varying Hubble parameter yields an almost
scale invariant primordial power spectrum.  Upon crossing the
horizon the phases of the quantum fluctuations freeze out and a
growing mode dominates the dynamics, i.e. the quantum fluctuations
become classical (see ref.\cite{liddle} and references therein).
Departures from scale invariance and gaussianity are determined by
the departures from flatness of the potential, namely by
derivatives of the potential with respect to the inflaton. These
derivatives can be combined into a hierarchy of dimensionless slow
roll parameters\cite{barrow} that allow an assessment of the
\emph{corrections} to the basic predictions of gaussianity and
scale invariance\cite{liddle}. The slow-roll approximation has
been recently cast  as a $1/N_{efolds}$ expansion\cite{n}, where
$N_{efolds}$ is  the number of efolds before the end of inflation
when modes of cosmological relevance today first crossed the Hubble
radius.

The basic scenario of inflation driven by a scalar field must be
interpreted as an  \emph{effective} field theory\cite{hector}
 resulting from integrating out  heavy degrees of
freedom. In particular, in the effective field theory description,
the \emph{classical} scalar potential that determines the dynamics
of the inflaton,   results from integrating out degrees of freedom
\emph{much heavier} than the scale of inflation.

Forthcoming observations have the potential of measuring the
inflationary potential at least within a span in field amplitude
corresponding to the 8-10 e-folds during which wavelengths of
cosmological relevance first cross the Hubble radius\cite{lidsey}. 
These observations will measure the {\it
full} inflaton potential including  {\it all} possible quantum
corrections and not just the classical (tree level) potential.
This possibility motivates us to assess the \emph{quantum}
corrections to the inflationary potential from fields lighter than
the inflaton, since in the effective field theory description, the
classical inflaton potential already includes contributions from
heavier fields. We focus on light fields since these can exhibit
infrared enhanced contributions to the effective potential as
discussed in ref.\cite{nuestros,nuestronor}.

Our goal is to obtain the \emph{effective} potential that includes
the one loop quantum  corrections from fields that are
\emph{light} during the relevant inflationary stage.

Our program of study focuses on the understanding of quantum
aspects of the basic inflationary paradigm. In previous studies we
addressed the decay of inflaton fluctuations\cite{nuestros} and
more recently\cite{nuestronor} we focused on the quantum
corrections to the equations of motion of the inflaton and the
scalar fluctuations during slow roll inflation, from integrating
out not only the inflaton fluctuations but also the excitations
associated with another scalar field. Since the power spectra of
fields with masses $m \ll H$ are nearly scale invariant, strong
infrared enhancements appear as revealed in these
studies\cite{nuestros,nuestronor}. In addition, we find that a
particular combination of slow roll parameters which measures the
departure from scale invariance of the fluctuations provides a
natural infrared regularization.

The small parameter that determines the validity of inflation as an effective
\emph{quantum field theory} below the Planck scale is $H/M_{Pl}$
where $H$ is the Hubble parameter during inflation and therefore the
scale at which inflation occurs. The slow roll expansion is in a
very well defined sense an \emph{adiabatic} approximation since the
time evolution of the inflaton field is slow on the expansion scale.
Thus the small dimensionless ratio $H/M_{Pl}$, which is required for
the validity of an effective field theory (EFT) is logically
\emph{independent} from the small dimensionless combinations of
derivatives of the potential which ensure the validity of the
slow-roll expansion. Present data\cite{peiris} indicate a very small
amplitude of tensor perturbations which  is consistent with $H/M_{Pl}\ll 1$.

Therefore, in this article we will invoke \emph{two independent}
approximations, the effective field theory (EFT) and the slow roll
approximation. The former is defined in terms of an expansion in
the ratio $H/M_{Pl}$, whereas the latter corresponds to an
expansion in the (small) slow roll parameters which has recently
been identified with an  expansion in $1/N_{efolds}$\cite{n}.

It is important to highlight the main differences between slow roll
inflation and the post-inflationary stage. During slow roll
inflation the dynamics of the scalar field is slow on the time scale
of the expansion and consequently the change in the amplitude
of the inflaton is small and quantified by the slow roll parameters.
The slow roll approximation is indeed an \emph{adiabatic approximation}.
In striking contrast to this situation, during the post-inflationary stage of
reheating the scalar field undergoes rapid and large amplitude oscillations
that cannot be studied in a perturbative expansion\cite{reheatnuestro,ramsey}.

{\bf Brief summary of results:}
  we obtain the quantum corrections to the inflaton potential
up to one loop by including the contributions from scalar and
tensor perturbations of the metric as well as one light scalar and
one light fermion field coupled generically to the inflaton.
Therefore this study provides the most complete assessment of the
general backreaction problem up to one loop that includes not only
metric perturbations, but also the contributions from fluctuations
of other light fields with a generic treatment of both bosonic and
fermionic degrees of freedom. Motivated by an assessment of the
quantum fluctuations that \emph{could} be of observational
interest, we focus on studying the effective inflaton potential
during the cosmologically relevant stage of slow roll inflation.

\vspace{2mm}

Both light bosonic fields as well
as scalar density perturbations feature an infrared enhancement of
their quantum corrections which is regularized by slow roll
parameters. Fermionic contributions as expected do not feature any
infrared enhancement and neither does the graviton contribution to
the energy momentum tensor.
We find that in slow roll and for light bosonic and fermionic
fields there is a clean separation between the super and
subhorizon contributions to the quantum corrections from scalar
density metric and light bosonic field perturbations. For these
fields the superhorizon contribution is of zero order in slow
roll as a consequence of the infrared enhancement regularized by
slow roll parameters. The subhorizon contribution to the energy
momentum tensor from all the fields is completely determined by
the trace anomaly of minimally coupled scalars, gravitons and
fermionic fields.  We find the one loop effective potential to be
\be
V_{eff}(\Phi_0) = V(\Phi_0)\Bigg[1+ \frac{H^2_0}{3 \; (4 \pi)^2 \;
M^2_{Pl}}\Bigg( \frac{\eta_v-4\,\epsilon_v}{\eta_v-3 \,
\epsilon_v}+\frac{3\,\eta_\sigma}{\eta_\sigma-\epsilon_v}+\mathcal{T}\Bigg)
  \Bigg] \label{Veffinsum}
\ee
  \noindent where $V(\Phi_0)$ is the \emph{classical} inflaton
  potential, $\eta_v, \epsilon_v, \eta_\sigma $ slow-roll parameters
and $ \mathcal{T} = \mathcal{T}_{\Phi}+ \mathcal{T}_s+\mathcal{T}_t
 +\mathcal{T}_{\Psi} =-\frac{2903}{20} =- 145.15$ is the total trace anomaly
from the scalar metric, tensor, light scalar and fermion contributions.

  The terms that feature ratios of slow roll parameters arise from
  superhorizon contributions from curvature and scalar field
  perturbations. The last term in eq.(\ref{Veffinsum}) is
independent of slow-roll parameters and is completely determined by the trace
anomalies of the different fields. It is the hallmark of the subhorizon
contributions.

In the case when the mass of the light bosonic scalar field is
much smaller than the mass of the inflaton fluctuations, we find
the following result for the scalar curvature and tensor
fluctuations including the one-loop quantum corrections, \bea &&
|{\Delta}_{k,eff}^{(S)}|^2 = |{\Delta}_{k}^{(S)}|^2 \left\{ 1+
\frac23 \left(\frac{H_0}{4 \; \pi \; M_{Pl}}\right)^2
\left[1+\frac{\frac38 \; r \; (n_s - 1) + 2 \; \frac{dn_s}{d \ln
k}}{(n_s - 1)^2} + \frac{2903}{40} \right] \right\} \cr \cr &&
|{\Delta}_{k,eff}^{(T)}|^2 =|{\Delta}_{k}^{(T)}|^2  \left\{ 1
-\frac13 \left(\frac{H_0}{4 \; \pi \; M_{Pl}}\right)^2
\left[-1+\frac18 \; \frac{r}{n_s - 1}+ \frac{2903}{20}
\right]\right\} \; , \cr \cr &&
r_{eff} \equiv \frac{|{\Delta}_{k,eff}^{(T)}|^2}{|{\Delta}_{k,eff}^{(S)}|^2}
= r \;  \left\{ 1-\frac13 \left(\frac{H_0}{4 \; \pi \; M_{Pl}}\right)^2
\left[1+\frac{\frac38 \; r \; (n_s - 1) +
\frac{dn_s}{d \ln k}}{(n_s - 1)^2} + \frac{8709}{20} \right] \right\} \; .
\eea 
The quantum corrections turn out to {\bf
enhance} the scalar curvature fluctuations and to {\bf reduce} the
tensor fluctuations as well as their ratio $r$. The quantum
corrections are always small, of the order $
\left(\frac{H_0}{M_{Pl}}\right)^2 $, but it is interesting to see
that these quantum effects are dominated by the trace anomalies
and they correct both scalar and tensor fluctuations in a definite
direction. Moreover, it is the tensor part of the trace anomaly
which numerically yields the largest contribution.

\bigskip

Quantum trace (conformal) anomalies of the energy momentum tensor
in gravitational fields constitute an important aspect of quantum
field theory in curved backgrounds, (see for example \cite{BD} and
references therein). In black hole backgrounds they are
related to the Hawking radiation. It is interesting to see here
that the trace anomalies appear in a relevant cosmological problem
and dominate the quantum corrections to the primordial spectrum of
curvature and tensor fluctuations.

\bigskip

  In section II we compute the effective potential including  scalars,
gravitons and fermionic fields, in section III we present the
quantum corrections to scalar curvature and tensor fluctuations,
and in section IV we present our conclusions.

\section{The effective potential}

In our recent calculation of the quantum corrections to the effective
potential\cite{nuestros}  different expansions appear: the expansion
in the effective field theory ratio $H_0/M_{Pl}$ where $H_0$ is the Hubble
parameter during the relevant stage of inflation, and the expansion in slow
roll parameters. These expansions are logically different: the
slow roll expansion is an \emph{adiabatic} expansion in the sense
that the dynamics of the inflaton is slower than the universe expansion,
while the (dimensionless) interaction vertices and the loop expansion are
determined by the effective field theory parameter $H_0/M_{Pl}$\cite{nuestros}.

During slow roll inflation, the dynamics of the scale factor and the
inflaton are determined by the following set of (semi) classical
equations of motion
\bea
&&H_0^2 = \frac{1}{3 \; M^2_{Pl}}\left[\frac{1}{2} \; (\dot{\Phi}_0)^2+
V(\Phi_0)\right]\;, \label{hub} \\ \label{claseq}
&&{\ddot \Phi}_0+3 \; H_0\,\dot{\Phi}_0+V'(\Phi_0) =0 \;.
\eea
where $ M_{Pl} =  1/\sqrt{8 \, \pi \; G} = 2.4 \; 10^{18}$GeV.
Slow roll inflation is tantamount to the statement that the dynamics
of the expectation value of the scalar field $\Phi_0$ is slow on the
scale of the cosmological expansion. The slow roll approximation is
indeed an \emph{adiabatic} approximation in terms of a hierarchy of
small dimensionless quantities related to the derivatives of the
inflaton potential. Some\cite{barrow,liddle} of these
slow roll parameters are given by\footnote{We follow the definitions
of $\xi_V;\sigma_V$ in ref.\cite{peiris}. ($\xi_V;\sigma_V$ are
called  $\xi^2_V;\sigma^3_V$, respectively, in\cite{barrow}).}
\bea
&&\epsilon_V  = \frac{M^2_{Pl}}{2} \;
\left[\frac{V^{'}(\Phi_0)}{V(\Phi_0)} \right]^2  \quad , \quad
\eta_V   = M^2_{Pl}  \; \frac{V^{''}(\Phi_0)}{V(\Phi_0)}\, ,
\label{etav} \\
&& \xi_V = M^4_{Pl} \; \frac{V'(\Phi_0) \;
V^{'''}(\Phi_0)}{V^2(\Phi_0)}  \quad , \quad  \sigma_V = M^6_{Pl}\;
\frac{\left[V^{'}(\Phi_0)\right]^2\,V^{(IV)}(\Phi_0)}{V^3(\Phi_0)}\;.
\nonumber \; 
\eea The slow roll approximation\cite{barrow,liddle,lidsey}
corresponds to $\epsilon_V \sim \eta_V \ll 1$  with the hierarchy
$\xi_V \sim \mathcal{O}(\epsilon^2_V)~;~\sigma_V \sim
\mathcal{O}(\epsilon^3_V)$, namely $\epsilon_V$ and $\eta_V$ are
first order in slow roll, $\xi_V$ second order in slow roll, etc.
Recently a correspondence between the slow roll expansion and an
expansion in $1/N_{efolds}$ has been established\cite{n} with
$\epsilon_V, \eta_V \sim 1/N_{efolds}~;~\xi_V \sim
1/N^2_{efolds}~;~\sigma_V \sim 1/N^3_{efolds}$,  etc.

During slow roll inflation the equation of motion (\ref{hub})-(\ref{claseq})
are approximated by
\bea
&&\dot{\Phi}_0 = -\frac{V'(\Phi_0)}{3 \; H_0}+ \textrm{higher order
in slow roll}\,, \label{slo} \cr \cr
\label{Fried}&& H^2_0 = \frac{V(\Phi_0)}{3 \; M^2_{Pl}}\Bigg[1+
\frac{\epsilon_V}{3}+\mathcal{O}(\epsilon^2_V,\epsilon_V\eta_V) \Bigg] \; ,
\eea
\noindent The scale factor is given by
\be\label{aoft}
C(\eta) = -\frac{1}{H \; \eta \; (1-\epsilon_V)} \; .
\ee
In the effective field theory interpretation of inflation, the
\emph{classical} inflaton potential $V(\Phi)$ should be understood
to include the contribution from  integrating out fields with
masses\emph{ much larger} than $H_0$. Our goal is to obtain the
one loop quantum corrections from \emph{fields that are light
during inflation}. Therefore we consider that the inflaton is
coupled to a light scalar field $\sigma$ and to Fermi fields
with a generic Yukawa-type coupling. We take the fermions to be Dirac fields
but it is straightforward to generalize to Weyl or Majorana
fermions. We also include the contribution to the effective
potential from scalar and tensor metric perturbations, thereby
considering their \emph{backreaction} up to one loop.

The Lagrangian density is taken to be
\be
\mathcal{L} = \sqrt{-g} \; \Bigg\{ \frac12 \;  \dot{\varphi}^2 -
\left(\frac{\vec{\nabla}\varphi}{2\,a}\right)^2-V(\varphi)+\frac12 \;
\dot{\sigma}^2 - \left(\frac{\vec{\nabla}\sigma}{2\,a}\right)^2 -
\frac12 \;  m^2_\sigma \;  \sigma^2 -G(\varphi)  \;  \sigma^2+
\overline{\Psi}\Big[i\,\gamma^\mu \;  \mathcal{D}_\mu \Psi -m_f -
Y(\varphi)\Big]\Psi \Bigg\} \label{lagrangian}
\ee
\noindent where $G(\Phi)$ and $Y(\Phi)$ are generic interaction terms
between the inflaton and the scalar and fermionic fields.
Obviously this Lagrangian can be further generalized to include a
multiplet of scalar and fermionic fields and such case can be
analyzed as a straightforward generalization. For simplicity we
consider one bosonic and one fermionic Dirac field.

The Dirac $\gamma^\mu$ are the curved space-time $\gamma$ matrices
and the fermionic covariant derivative is given
by\cite{weinberg,BD,duncan,casta}
\bea
\mathcal{D}_\mu & = &  \partial_\mu + \frac{1}{8} \;
[\gamma^c,\gamma^d] \;  V^\nu_c  \; \left(D_\mu V_{d \nu} \right)
\cr \cr
D_\mu V_{d \nu} & = & \partial_\mu V_{d \nu} -\Gamma^\lambda_{\mu
\nu} \;  V_{d \lambda} \nonumber
\eea
\noindent where the vierbein field is defined as
$$
g^{\mu\,\nu} = V^\mu_a \;  V^\nu_b \;  \eta^{a b} \; ,
$$
\noindent $\eta_{a b}$ is the Minkowski space-time metric and
the curved space-time  matrices $\gamma^\mu$ are given in terms of
the Minkowski space-time ones $\gamma^a$  by (greek indices refer to
curved space time coordinates and latin indices to the local
Minkowski space time coordinates)
$$
\gamma^\mu = \gamma^a V^\mu_a \quad , \quad
\{\gamma^\mu,\gamma^\nu\}=2 \; g^{\mu \nu}  \; .
$$
We will consider that the light scalar field $\sigma$ has vanishing
expectation value at all times, therefore inflationary dynamics is
driven by one single scalar field, the inflaton $\phi$. We now
separate the homogeneous expectation value of the inflaton field
from its quantum fluctuations as usual by writing
$$
\varphi(\vec{x},t) = \Phi_0(t) +\delta\varphi (\vec{x},t) \; .
$$
We will consider the contributions from the quadratic fluctuations
to the energy momentum tensor. There are \emph{four} distinct
contributions: i) scalar metric (density) perturbations, ii) tensor
(gravitational waves) perturbations, iii) fluctuations of the light
bosonic scalar field $\sigma$, iv) fluctuations of the light
fermionic field $\Psi$.

Fluctuations in the metric are studied as
usual \cite{mukha,mukhanov,giova,hu,riotto}. Writing the metric as
$$
g_{\mu\nu}= g^0_{\mu\nu}+\delta^s g_{\mu\nu}+\delta^t g_{\mu\nu}
$$
\noindent where $g^0_{\mu\nu}$ is the spatially flat FRW background
metric which in conformal time is given by
$$
g^0_{\mu\nu}= C^2(\eta) \;  \eta_{\mu\nu}
\quad , \quad C(\eta)\equiv a(t(\eta))
$$
\noindent and $\eta_{\mu\nu}=\textrm{diag}(1,-1,-1,-1)$ is the flat
Minkowski space-time metric.  $\delta^{s,t} g_{\mu\nu}$ correspond
to the scalar and tensor perturbations respectively, and we neglect
vector perturbations. In longitudinal gauge
\bea
\delta^{s} g_{00} & = & C^2(\eta) \; 2  \; \phi \cr \cr
\delta^{s} g_{ij} & = & C^2(\eta) \;  2 \;  \psi \;  \delta_{ij}
\label{curvpot}\\
\delta^{t} g_{ij} & = & -C^2(\eta) \;  h_{ij} \nonumber
\eea
\noindent where $h_{ij}$ is transverse and
traceless and we neglect vector modes since they are not
generated in single field inflation\cite{mukha,mukhanov,giova,hu,riotto}.

Gauge invariant variables associated with the fluctuations of the
scalar field and the potentials $\phi,\psi$ are constructed
explicitly in ref.\cite{mukhanov} where the reader can find their
expressions. Expanding up to quadratic order in the
scalar fields, fermionic fields and metric perturbations the part
of the Lagrangian density that is quadratic in these fields is
given by
$$
\mathcal{L}_Q =\mathcal{L}_s[\delta\varphi^{gi},\phi^{gi},\psi^{gi}]+
\mathcal{L}_t[h]+\mathcal{L}_\sigma[\sigma]+
\mathcal{L}_\Psi[\overline{\Psi},\Psi] \; ,
$$
\noindent where
\bea
&&\mathcal{L}_t[h] = \frac{M^2_{Pl}}{8} \; C^2(\eta) \;
\partial_\alpha h^j_i  \; \partial_\beta h^i_j \;  \eta^{\alpha \beta} \; ,
\cr \cr
&&
\mathcal{L}_\sigma[\sigma]=
 C^4(\eta) \; \Bigg\{\frac{1}{2} \;
\left(\frac{\sigma'}{C}\right)^2-\frac{1}{2} \; \left(\frac{\nabla
\sigma}{ C}\right)^2 -\frac{1}{2} \; M^2_\sigma[\Phi_0]\;
\sigma^2\Bigg\}  \; , \cr \cr
&& 
\mathcal{L}_\Psi[\overline{\Psi},\Psi]= \overline{\Psi}\Big[i \;
\gamma^\mu  \; \mathcal{D}_\mu \Psi -M_\Psi[\Phi_0]\Big]\Psi \; ,
\nonumber \eea \noindent where the prime stands for derivatives
with respect to conformal time and the labels (gi) refer to gauge
invariant quantities\cite{mukhanov}. The explicit expression for
$\mathcal{L}[\delta\varphi^{gi},\phi^{gi},\psi^{gi}]$ is given in
eq. (10.68) in ref.\cite{mukhanov}. The effective masses for the
bosonic and fermionic fields are given by \bea M^2_\sigma[\Phi_0]
& = & m^2_\sigma + G(\Phi_0) \label{sigmamass}\\ \cr
M_\Psi[\Phi_0] & = & m_f+Y(\Phi_0) \; . \nonumber \eea We will
focus on the study of the quantum corrections to the Friedmann
equation, for the case in which both the scalar and fermionic
fields are light in the sense that during slow roll inflation, \be
M_\sigma[\Phi_0], \; M_\Psi[\Phi_0] \ll H_0 \; , \ee \noindent at
least during the cosmologically relevant stage corresponding to
the 50 or so e-folds before the end of inflation.

In conformal time the vierbeins $V^\mu_a$ are particularly simple
\be
V^\mu_a = C(\eta) \; \delta^\mu_a
\ee
\noindent and the Dirac Lagrangian density simplifies to the
following expression
\be \label{ecdi}
\sqrt{-g} \; \overline{\Psi}\Bigg(i \; \gamma^\mu \;  \mathcal{D}_\mu
\Psi -M_\Psi[\Phi_0]\Bigg)\Psi  =
C^{\frac{3}{2}}\overline{\Psi} \;  \Bigg[i \;
{\not\!{\partial}}-M_\Psi[\Phi_0] \; C(\eta) \Bigg]
\left(C^{\frac{3}{2}}{\Psi}\right)
\ee
\noindent where $i {\not\!{\partial}}$ is the usual Dirac
differential operator in Minkowski space-time in terms of flat
space time $\gamma$ matrices.

>From the quadratic Lagrangian given above the  quadratic quantum
fluctuations to the energy momentum tensor can be extracted.

The effective potential is identified with $\langle T^0_0 \rangle$
in a spatially translational invariant state in which the
expectation value of the inflaton field is $\Phi_0$. During slow
roll inflation the expectation value $\Phi_0$ evolves very slowly
in time, the slow roll approximation is indeed an adiabatic
approximation, which justifies treating $\Phi_0$ as a constant in
order to obtain the effective potential. The time variation of
$\Phi_0$ only contributes to higher order corrections in
slow-roll. This is standard in \emph{any} calculation of an
effective potential. The energy momentum tensor is computed in the
FRW inflationary background determined by the \emph{classical}
inflationary potential $V(\Phi_0)$, and the slow roll parameters
are also explicit functions of $\Phi_0$. Therefore the energy
momentum tensor depends \emph{implicitly} on $\Phi_0$ through the
background and \emph{explicitly} on the masses for the light
bosonic and fermionic fields given above.

Therefore the effective potential is given by \be V_{eff}(\Phi_0)
= V(\Phi_0)+ \delta V(\Phi_0)\label{Veff} \ee \noindent where \be
\delta V(\Phi_0)= \langle T^{0}_{0}[\Phi_0] \rangle_s + \langle
T^{0}_{0}[\Phi_0] \rangle_t + \langle T^{0}_{0}[\Phi_0]
\rangle_\sigma +\langle T^{0}_{0}[\Phi_0]
\rangle_\Psi\label{dVeff} \ee \noindent $(s,t,\sigma,\Psi)$
correspond to the energy momentum tensors of the quadratic
fluctuations of the scalar metric, tensor (gravitational waves),
light boson field $\sigma$ and light fermion field $\Psi$
fluctuations respectively. Since these are the expectation values
of a quadratic energy momentum tensor, $\delta V(\Phi_0)$
corresponds to the \emph{one loop correction} to the effective
potential.

\subsection{Light scalar fields}

We begin by analyzing the contribution to the effective potential
from the light bosonic scalar field $\sigma$ because this study
highlights the main  aspects which are relevant in the case
of scalar metric (density) perturbations.

The bosonic  Heisenberg field operators are expanded as follows
\be
\sigma(\vec{x},\eta) = \frac{1}{C(\eta) \; \sqrt{\Omega}}
\sum_{\vec{k}}\,  e^{i \vec{k}\cdot \vec{x}}\left[a_{\sigma,\vec{k}}
\, S_\sigma(k,\eta)+ a^\dagger_{\sigma,\vec{k}} \,S^*_\sigma(k,\eta)
\right] \label{sigm}
\ee
\noindent where $\Omega$ is the spatial volume.

During slow roll inflation the effective mass of the $\sigma$
field is given by eq. (\ref{sigmamass}), just as for the inflaton
fluctuation. It is convenient to introduce a parameter
$\eta_\sigma$ defined to be \be\label{etasig} \eta_\sigma =
\frac{M^2_\sigma[\Phi_0]}{3 \;  H^2_0 } \; . \ee Hence, the
statement that the $\sigma$ field is light corresponds to the
condition \be \eta_\sigma \ll 1 \label{light} \; . \ee This
dimensionless parameter plays the same role for the $\sigma$ field
as the parameter $\eta_V$ given by eq. (\ref{etav}) does for the
inflaton fluctuation.

The mode functions $S_\sigma(k,\eta)$ in eq.
(\ref{sigm}) obey the following equations up to quadratic
order\cite{nuestronor}
$$
 S^{''}_{\sigma}(k,\eta)+ \left[k^2 + M^2_{\sigma}(\Phi_0) \;  C^2(\eta)-
\frac{C^{''}(\eta)}{C(\eta)} \right]S_{\sigma}(k,\eta)=0 \,.
$$
Using the slow roll expressions eq.(\ref{aoft}) and in terms of
$\eta_\sigma$, these mode equations  become
$$
S^{''}_{\sigma}(k,\eta)+\left[k^2-
\frac{\nu^2_{\sigma}-\frac{1}{4}}{\eta^2} \right]S_{\sigma}(k,\eta)
= 0 ~~;~~ \nu_\sigma = \frac{3}{2}+\epsilon_V-\eta_{\sigma} +
\mathcal{O}(\epsilon^2_V,\eta^2_{\sigma},\eta^2_V,\epsilon_V
\eta_V) \; .
$$
During slow roll inflation $\Phi_0$ is approximately constant, and
the slow roll expansion is an \emph{adiabatic} expansion. As usual
in the slow roll approximation, the above equation for the mode
functions is solved by assuming that $\Phi_0$, hence $\nu_\sigma$
are \emph{constant}. This is also the same type of approximation
entailed in \emph{every} calculation of the effective potential.
Therefore during slow roll, the solution of the mode functions above
are
$$
S_{\sigma}(k,\eta) = \frac{1}{2} \; \sqrt{-\pi\eta}
\; e^{i\frac{\pi}{2}(\nu_\sigma+\frac{1}{2})} \;
H^{(1)}_{\nu_\sigma}(-k\eta) \;  .
$$
This choice of mode functions  defines the Bunch-Davis vacuum, which
obeys $a_{\vec{k}}|0>_{BD}=0$. It is important to highlight that
there is no unique choice of vacuum or initial state, a recognition
that has received considerable attention in the literature, see for
example\cite{inistate,mottola} and references therein. In this study
we focus on  Bunch-Davis initial conditions since this has been the
standard choice to study the power spectra and metric perturbations,
hence we can compare our results to the standard ones in the
literature,  postponing for further study the assessment of
different initial states.

The contribution to the effective potential from the light scalar
field $\sigma$ is given by
$$ 
\langle T^0_{0} \rangle_\sigma =
\frac{1}{2}\,\Bigg\langle\dot{\sigma}^2+ \left(\frac{\nabla
\sigma}{C(\eta)}\right)^2+M^2_\sigma[\Phi_0] \;  \sigma^2
\Bigg\rangle \; ,
$$
\noindent where the dot stands for derivative with respect to cosmic
time. The expectation values are in the Bunch-Davis vacuum state and
yield the following contributions
\bea
\frac{1}{2} \; \big\langle\left(\dot{\sigma}\right)^2 \big\rangle
& = & \frac{H^4_0}{16\pi}\int_0^\infty \frac{dz}{z} \;  z^2 \;
\Big|\frac{d}{dz} \left[z^\frac{3}{2}H^{(1)}_{\nu_\sigma}(z) \right]
\Big|^2 \label{sigdot2} \\
\frac{1}{2}\,\big\langle\ \left(\frac{\nabla \sigma}{C^2(\eta)}\right)^2
\big \rangle & = & \frac{H^4_0}{16\pi}\int_0^\infty \frac{dz}{z} \;  z^5 \;
\Big|H^{(1)}_{\nu_\sigma}(z)  \Big|^2 \label{gradsig2} \\
\frac{M^2_\sigma[\Phi_0]}{2} \; \big \langle \sigma^2(\vec{x},t) \big
\rangle & = & \frac{3\,H^2_0\,\eta_\sigma}{2} \int_0^\infty
\frac{dk}{k} \;  \mathcal{P}_\sigma(k,t) \label{sig2} \; ,
\eea
\noindent where $\mathcal{P}_\sigma(k,t)$ is the power spectrum of
the $\sigma$ field, which in terms of the spatial Fourier transform
of the field $\sigma_{\vec{k}}(t)$ is given by
$$
\mathcal{P}_\sigma(k,t) = \frac{k^3}{2\pi^2} \; \big
\langle \left|\sigma^2_{\vec{k}}(t)\right| \big \rangle =
\frac{H^2_0}{8\pi} \;  (-k\eta)^3 \;
\left|H^{(1)}_{\nu_\sigma}(-k\eta)\right|^2  \; .
$$
For a light scalar field during slow roll the power spectrum of the
scalar field $\sigma$ is nearly scale invariant and the index
$\nu_\sigma \sim 3/2$. In the exact scale invariant case $\nu_\sigma = 3/2$,
$$
z^3 \left|H^{(1)}_\frac{3}{2}(z)\right|^2 =
\frac{2}{\pi} [1 + z^2]
$$
\noindent and  the integral of the power spectrum in eq.
(\ref{sig2}) not only features logarithmic and quadratic
\emph{ultraviolet} divergences but also a logarithmic
\emph{infrared} divergence. During slow roll and for a light but
massive scalar field the quantity
$$
\Delta_\sigma = \frac{3}{2}-\nu_\sigma = \eta_\sigma-\epsilon_V +
\mathcal{O}(\epsilon^2_V,\eta^2_\sigma,\epsilon_V\,\eta_\sigma) \; ,
\ll 1
$$
\noindent is a measure of the departure from scale invariance and
provides a natural \emph{infrared regulator}. We note that the
contribution from eq. (\ref{sig2}) to the effective potential,
which can be written as
$$
\frac{3\,H^4_0\,\eta_\sigma}{16 \; \pi} \int_0^\infty
\frac{dz}{z} \; z^3 \;  \left|H^{(1)}_{\nu_\sigma}(z)\right|^2  \; ,
$$
\noindent is \emph{formally} smaller than the contributions from
eqs.(\ref{sigdot2})-(\ref{gradsig2}) by a factor $\eta_\sigma \ll 1$.
However, the logarithmic infrared divergence in the exact scale
invariant case, leads to a single \emph{pole} in the variable
$\Delta_\sigma$ as described in detail in refs\cite{nuestros,nuestronor}.
To see this feature in detail, it proves convenient to separate
the infrared contribution by writing the integral above in the
following form
$$
\int_0^\infty \frac{dz}{z} \; z^3 \;
\left|H^{(1)}_{\nu_\sigma}(z)\right|^2 = \int_0^{\mu_p}
\frac{dz}{z} \; z^3 \;  \left|H^{(1)}_{\nu_\sigma}(z)\right|^2 +
\int_{\mu_p}^\infty \frac{dz}{z} \; z^3 \;
\left|H^{(1)}_{\nu_\sigma}(z)\right|^2  \; .
$$
In the first integral we obtain the leading order contribution in
the slow roll expansion, namely the pole in $\Delta_\sigma$, by using the
small argument limit of the Hankel functions
$$
z^3 \, \left|H^{(1)}_{\nu_{\sigma}} (z)\right|^2 \buildrel{z \to 0}\over=\left[
\frac{2^{\nu_\sigma} \; \Gamma(\nu_\sigma)}{\pi} \right]^2 \; z^{2 \,
\Delta_\sigma}
$$
\noindent which yields
$$
\int^{\mu_p}_0 \frac{dz}{z} \; z^3 \,
\left|H^{(1)}_{\nu_{\sigma}} (z)\right|^2 = \frac{2}{\pi}\left[\frac{1}{2 \,
\Delta_\sigma}+ \frac{\mu^2_p}{2} + \gamma - 2 + \ln(2 \; \mu_p)
+\mathcal{O}(\Delta_\sigma)\right]\,,
$$
\noindent
In the second integral for small but fixed $\mu_p$, we can safely
set $\Delta_\sigma=0$ and by introducing an upper momentum (ultraviolet)
cutoff $\Lambda_p$, we finally find
$$
\int_0^{\Lambda_p} \frac{dz}{z} \; z^3 \;
\left|H^{(1)}_{\nu_\sigma}(z)\right|^2 = \frac{1}{\pi}
\left[\frac{1}{\Delta_\sigma} + {\Lambda_p}^2 + \ln \Lambda_p^2
 + 2 \, \gamma - 4 + \mathcal{O}(\Delta_\sigma) \right]
$$
 The simple pole in $\Delta_\sigma$ reflects the infrared enhancement
 arising from a nearly scale invariant power spectrum. While the
 terms that depend on $\Lambda_p$ are of purely ultraviolet origin
and correspond to the specific regularization scheme, the
 simple pole in $\Delta_\sigma$ originates in the \emph{infrared} behavior
 and is therefore independent of the regularization scheme. A
 covariant regularization of the expectation value $\langle
 \sigma^2(\vec x,t)\rangle$ will yield a result which
features a simple pole in $\Delta_\sigma$ plus terms which are
ultraviolet finite and regular in the limit $\Delta_\sigma
\rightarrow 0$. Such regular terms
 yield a contribution  $\mathcal{O}(H^4 \; \eta_\sigma)$
to eq.(\ref{sig2}) and are subleading in the limit of light scalar fields
because they do not feature a denominator $\Delta_\sigma$.

 Therefore, to leading order in the slow roll expansion and in
 $\eta_\sigma \ll 1$, the contribution from eq.(\ref{sig2}) is given by,
$$
\frac{M^2_\sigma[\Phi_0]}{2} \; \big \langle \sigma^2(\vec{x},t) \big
\rangle =
 \frac{3\,H^4_0}{(4 \; \pi)^2} \; \frac{\eta_\sigma}{ \eta_\sigma-\epsilon_V}+
\mathrm{subleading  ~in ~ slow ~ roll}.
$$
 In the first two contributions given by eqs.(\ref{sigdot2})-(\ref{gradsig2})
 extra powers of momentum arising either from the time or spatial
 derivatives, prevent the logarithmic infrared enhancements. These
 terms are infrared finite in the limit $\Delta_\sigma \rightarrow 0$ and
 their leading contribution during slow roll can be obtained by
 simply setting $\nu_\sigma = 3/2$ in these integrals, which feature
 quartic, quadratic and logarithmic ultraviolet divergences. A
 covariant  renormalization of these two terms will lead to an
 ultraviolet and an infrared
 finite contribution to the energy momentum tensor of
 $\mathcal{O}(H^4_0)$, respectively.
For the term given by eq.(\ref{sig2}), the
 infrared contribution that yields the pole in $\Delta_\sigma$ compensates
 for the  $\eta_\sigma \ll 1 $ in the numerator, after
 renormalization of the ultraviolet divergence, the ultraviolet and
 infrared finite contributions to this term will yield a
 contribution to the energy momentum tensor of order
 $\mathcal{O}(H^4_0 \; \eta_\sigma)$, without the small denominator,
and therefore subleading. This analysis
 indicates that the leading order contributions to the energy
 momentum tensor for light scalar fields is determined by the
 infrared pole $\sim 1/\Delta_\sigma$ from eq.(\ref{sig2}) and the fully
 renormalized contributions from (\ref{sigdot2})-(\ref{gradsig2}),
 namely to leading order in slow roll and $\eta_\sigma$
 \be
\langle T^0_0 \rangle_\sigma =
 \frac{3\,H^4_0}{(4 \; \pi)^2}\frac{\eta_\sigma}{ \frac{3}{2}-\nu_\sigma}+
\frac{1}{2}\,\Bigg\langle\dot{\sigma}^2+ \left(\frac{\nabla
\sigma}{C(\eta)}\right)^2 \Bigg\rangle_{ren} \label{T00siglead}
\ee
In the expression above we have displayed explicitly the pole at
$3/2-\nu_\sigma = \eta_\sigma-\epsilon_V$.

In calculating the second term (renormalized expectation value)
to leading order in eq.(\ref{T00siglead}) we can set to zero the slow
roll parameters $\epsilon_V, \eta_V$ as well as the mass of the
light scalar, namely $\eta_\sigma=0$. Hence, to leading order,
the second term is identified with the $00$ component of the
renormalized energy momentum tensor for a free massless minimally
coupled scalar field in exact de Sitter space time. Therefore we can
extract this term from the known result for the renormalized
energy momentum tensor for a minimally coupled free scalar boson of
mass $m_\sigma$ in de Sitter space time with a Hubble constant
$H_0$ given by\cite{BD,fordbunch,sanchez}
\bea
\langle T_{\mu \nu}\rangle_{ren} &=& \frac{g_{\mu
\nu}}{(4 \, \pi)^2}\Bigg\{m^2_\sigma \; H^2_0
\left(1-\frac{m^2_\sigma}{2 \, H^2_0}\right)\left[-\psi\left(\frac{3}{2}+
\nu\right)-\psi\left(\frac{3}{2}-\nu\right)+\ln\frac{m^2_\sigma}{H^2_0} \right]+
\frac{2}{3} \; m^2_\sigma  \; H^2_0-\frac{29}{30} \; H^4_0 \Bigg\} \; , \cr \cr
 \nu &\equiv&  \sqrt{\frac{9}{4}-\frac{m^2_\sigma}{H^2_0}}\label{TmunudS} \; .
\eea
where $ \psi(z) $ stands for the digamma function.
This expression corrects a factor of two in
ref.\cite{BD,dowker}. In eq. (6.177) in \cite{BD} the
D'Alambertian acting on $G^{1}(x,x')$ was neglected. However, in
computing this term, the D'Alambertian must be calculated
\emph{before} taking the coincidence limit. Using the equation of
motion yields the extra factor 2 and the expression eq.(\ref{TmunudS}).
This result  eq.(\ref{TmunudS}) for the renormalized energy momentum
tensor was obtained by several different methods: covariant point splitting,
zeta-function and Schwinger's proper time
regularizations\cite{BD,dowker}.

The simple pole at $ \nu=3/2 $ manifest in eq.(\ref{TmunudS})
coincides precisely with the similar simple pole in eq.
(\ref{T00siglead}) as can be gleaned by recognizing that
$m^2_\sigma = 3 \; H^2 \;  \eta_\sigma$ as stated by
eq.(\ref{etasig}). This pole originates in the term $m_\sigma^2 \;
<\sigma^2>$, which features an infrared divergence in the scaling
limit $\nu_\sigma=3/2$. All the terms that contribute to the
energy momentum tensor with space-time derivatives are infrared
finite in this limit. Therefore, from the energy momentum tensor
eq.(\ref{TmunudS}) we can extract straightforwardly the leading
contribution to the renormalized expectation value in
eq.(\ref{T00siglead}) in the limit $H_0 \gg m_\sigma$, and
neglecting the slow roll corrections to the scale factor. It is
given by the last term in the bracket in eq. (\ref{TmunudS}).
Hence, we find the leading order contribution \be \langle T^0_0
\rangle_\sigma =
 \frac{H^4_0}{(4 \; \pi)^2}\left[\frac{3\,\eta_\sigma}{
 \eta_\sigma-\epsilon_V}-\frac{29}{30}+
\mathcal{O}(\epsilon_V,\eta_\sigma,\eta_V)\right]\label{Tsiglea}
\ee
The last term is completely determined by the trace
anomaly\cite{BD,sanchez,fordbunch,duff,dowker,hartle,fujikawa}
which is in turn determined by the short distance correlation
function of the field and the background geometry.

Therefore, we emphasize that in the slow roll approximation there
is a clean and unambiguous separation between the contribution from
superhorizon modes, which give rise to simple poles in slow roll
parameters and that of subhorizon modes whose leading contribution
is determined by the trace anomaly and the short distance behavior
of the field.

\subsection{Scalar metric perturbations}

The gauge invariant energy momentum tensor for quadratic scalar
metric fluctuations has been obtained in ref.\cite{abramo} where
the reader is referred to for details. In longitudinal gauge and
in cosmic time it is given  by
\bea
\langle T^0_0 \rangle_s = && M^2_{Pl} \Bigg[12 \; H_0 \;  \langle \phi
\dot{\phi} \rangle - 3 \;  \langle (\dot{\phi})^2 \rangle +
\frac{9}{C^2(\eta)} \; \langle (\nabla \phi)^2\rangle \Bigg] \nonumber \\
& & + \frac{1}{2} \; \langle (\dot{\delta \varphi})^2\rangle + \frac{\langle
(\nabla \delta \varphi)^2\rangle}{2\,C^2(\eta)} + \frac{V''(\Phi_0)}{2} \;
\langle(\delta \varphi)^2\rangle + 2 \;  V'(\Phi_0) \;
\langle \phi \,\delta \varphi
\rangle \label{T00s}
\eea
\noindent where the condition $\phi = \psi$ valid in scalar field
inflation has been used, and the dots stand for derivatives with
respect to cosmic time.

In longitudinal gauge, the equations of motion in cosmic time for
the Fourier modes are\cite{mukhanov,riotto} \bea \label{phieq}
&&\ddot{\phi}_{\vec k}+ \left(H_0-2 \;
\frac{\ddot{\Phi}_0}{\dot{\Phi}_0}\right)\dot{\phi}_{\vec k}+
\left[2 \; \left(\dot{H}_0-2 \; H_0 \;
\frac{\ddot{\Phi}_0}{\dot{\Phi}_0}\right)+
\frac{k^2}{C^2(\eta)}\right]{\phi}_{\vec k}=0 \cr \cr &&
\label{delfieqn} \ddot{\delta \varphi}_{\vec k}+3 \; H \;
\dot{\delta \varphi}_{\vec
k}+\left[V''[\Phi_0]+\frac{k^2}{C^2(\eta)} \right]\delta
\varphi_{\vec k}+2 \;  V'[\Phi_0] \; \phi_{\vec k}- 4 \;
\dot{\Phi}_0 \; \dot{\phi}_{\vec k}=0  \; , \eea \noindent with
the constraint equation \be \label{constraint} \dot{\phi}_{\vec
k}+H_0 \; \phi_{\vec k}= \frac{1}{2M_{Pl}} \; \delta \varphi_{\vec
k} \; \dot{\Phi}_0  \; . \ee Just as in the case of the scalar
fields, we expect an infrared enhancement arising from
superhorizon modes, therefore, following ref.\cite{abramo}  we
split the contributions to the energy momentum tensor as those
from superhorizon modes, which will yield the infrared
enhancement, and the subhorizon modes for which we can set all
slow roll parameters to zero. Just as discussed above for the case
of the $\sigma$ field, since spatio-temporal derivatives bring
higher powers of the momenta, we can neglect all derivative terms
for the contribution from the superhorizon modes. Therefore, the
contribution from superhorizon modes which will reflect the
infrared enhancement is extracted from\cite{abramo} \be \langle
T^0_0 \rangle_{IR} \approx \frac{1}{2} \;  V''[\Phi_0] \; \langle
\left(\delta \varphi (\vec{x},t)\right)^2 \rangle + 2 \;
V'[\Phi_0] \;  \langle \phi(\vec{x},t)\,\delta \varphi(\vec{x},t)
\rangle  \; . \ee The analysis of the solution of eq.(\ref{phieq})
for superhorizon wavelengths in ref. \cite{mukhanov} shows that in
exact de Sitter space time $\phi_{\vec k} \sim \mathrm{constant}$,
hence it follows that during quasi-de Sitter slow roll inflation
for superhorizon modes \be \label{fipun} \dot{\phi}_{\vec k} \sim
(\mathrm{slow~roll}) \times H_0 \; \phi_{\vec k} \ee Therefore,
for superhorizon modes, the  constraint equation
(\ref{constraint}) yields \be \label{rela} \phi_{\vec k} = -\,
\frac{V'(\Phi_0)}{2 \; V(\Phi_0)}  \;  \delta \varphi_{\vec k} +
{\rm higher ~ orders ~ in ~ slow ~ roll} \; . \ee Inserting this
relation in eq.(\ref{delfieqn}) and consistently neglecting the
term $\dot{\phi}_{\vec k}$ according to eq.(\ref{fipun}), we find
the following equation of motion for the gauge invariant scalar
field fluctuation in longitudinal gauge \be \label{delfieqn2}
\ddot{\delta \varphi}_{\vec k}+3 \; H_0  \; \dot{\delta
\varphi}_{\vec k}+\left[\frac{k^2}{C^2(\eta)}+3 \;  H^2_0 \;
\,\eta_\delta \right]\delta \varphi_{\vec k}=0  \; , \ee \noindent
where we have used the definition of the slow roll parameters
$\epsilon_V; \; \eta_V$ given in eq.(\ref{etav}), and introduced
\be\label{etadelta} \eta_\delta \equiv \eta_V-2 \; \epsilon_V \ee
This is the equation of motion for a minimally coupled scalar
field with mass squared $3 \;  H^2_0  \; \eta_\delta$ and we can
use the results obtained in the case of the scalar field $\sigma$
above. The quantum field $\delta \varphi(\vec x,t) $ is expanded
as \be \delta \varphi(\vec{x},\eta) = \frac{1}{C(\eta) \;
\sqrt{\Omega}} \sum_{\vec{k}}\,  e^{i \vec{k}\cdot
\vec{x}}\left[a_{\delta,\vec{k}}
 \;  S_\delta(k,\eta)+ a^\dagger_{\delta,\vec{k}} \;  S^*_\delta(k,\eta)
\right] \label{deltaexp}  \; ,
\ee
\noindent where the mode functions are given by
\be \label{delmod} S_{\delta}(k,\eta) = \frac{1}{2} \; \sqrt{-\pi\eta}
\; e^{i\frac{\pi}{2}(\nu_\delta+\frac{1}{2})} \;
H^{(1)}_{\nu_\delta}(-k\eta)~~;~~ \nu_\delta =
\frac{3}{2}+\epsilon_V-\eta_\delta = \frac{3}{2}+3 \; \epsilon_V-\eta_V
 \; .
\ee
In this case, the slow roll quantity that regulates the infrared
behavior is $\Delta_\delta \equiv \eta_V-3 \; \epsilon_V$.

Again we choose the Bunch-Davies vacuum state annihilated by the
operators $a_{\delta,\vec{k}}$. Therefore, the  contribution to
$\langle T^0_{0}\rangle$ from superhorizon modes to lowest order
in slow roll is given by \be \langle T^0_{0} \rangle_{IR} = 3 \;
H^2_0 \left(\frac{\eta_V}{2}-2\,\epsilon_V\right)
\left[\int_0^\infty \frac{dk}{k} \mathcal{P}_\delta(k,\eta)
\right]_{IR} \ee \noindent where the power spectrum of scalar
fluctuations is given by \be\label{powspdel}
\mathcal{P}_\delta(k,\eta) = \frac{k^3}{2\pi^2} \; \big \langle
\left|\delta \varphi_{\vec{k}}(t)\right|^2 \big \rangle =
\frac{H^2_0}{8\pi} \; (-k\eta)^3 \;
\left|H^{(1)}_{\nu_\delta}(-k\eta)\right|^2 \ee \noindent and the
subscript $IR$ in the integral refers only to the infrared pole
contribution to $\Delta_\delta$. Repeating the analysis presented
in the case of the scalar field $\sigma$ above, we finally find
\be \langle T^0_0 \rangle_{IR} = \frac{3 \; H^4_0}{(4 \; \pi)^2}
\; \frac{\eta_V-4\,\epsilon_V}{\eta_V-3 \, \epsilon_V} +
\mathrm{subleading~in~slow~roll} \ee For subhorizon modes with
wavevectors $k \gg a(t) \;  H_0$, the solutions of the equation
(\ref{phieq}) are\cite{mukhanov} \be \label{apro} \phi_{\vec k}(t)
\approx e^{\pm i k \eta} \Rightarrow \dot{\phi}_{\vec k}(t) \sim
\frac{i \, k}{a(t)} \,{\phi}_{\vec k}(t) \ee For $k \gg a(t) \;
H_0$ the constraint equation (\ref{constraint}) entails
that\cite{abramo} \be \label{conss} \phi_{\vec k}(t) \approx
\frac{i \, a(t)}{2\,M_{Pl}\,k} \; \dot{\Phi}_0  \;
\delta\varphi_{\vec k}. \ee Replacing the expressions
eqs.(\ref{apro})-(\ref{conss}) in eq.(\ref{T00s}) yields that all
the terms featuring the gravitational potential $\phi$ are
suppressed with respect to those featuring the scalar field
fluctuation $\delta\varphi$ by powers of $H_0 \; a(t)/k \ll 1$ as
originally observed in ref.\cite{abramo}. Therefore the
contribution from subhorizon modes to $\langle T^0_{0s} \rangle$
is given by \be \langle T^0_{0s}\rangle_{UV} \simeq
\frac{1}{2}\langle (\dot{\delta \varphi})^2 \rangle +
\frac{\langle (\nabla \delta \varphi)^2\rangle}{2\,a^2} \ee
\noindent where we have also neglected the term with $V''[\Phi_0]
\sim 3 \; H^2_0 \; \eta_V $ since  $ k^2/a^2 \gg H^2_0 $ for
subhorizon modes. Therefore, to leading order in slow roll we find
the renormalized expectation value of $T_{00s}$ is given by \be
\label{T00sfin} \langle T^0_{0s}\rangle_{ren} \simeq \frac{3
H^4_0}{(4 \; \pi)^2} \frac{\eta_V-4\,\epsilon_V}{\eta_V-3 \,
\epsilon_V} + \frac{1}{2}\Bigg\langle \dot{\delta \varphi}^2 +
\left(\frac{\nabla \delta \varphi}{C(\eta)}\right)^2
\Bigg\rangle_{ren} \ee To obtain the renormalized expectation
value in eq.(\ref{T00sfin}) one can set all slow roll parameters
to zero to leading order and simply consider a massless scalar
field minimally coupled in de Sitter space time. This is precisely
what we have already calculated in the case of the scalar field
$\sigma$ above by using the known results in the literature for
the covariantly renormalized energy momentum tensor of a massive
minimally coupled field\cite{BD,fordbunch,sanchez,dowker}, and we
can just borrow the result from eq.(\ref{Tsiglea}). We find the
following final result to leading order in slow roll \be \langle
T^0_{0s} \rangle_{ren} =
 \frac{H^4_0}{(4 \; \pi)^2}\left[\frac{\eta_V-4\,\epsilon_V}{\eta_V-3 \,
\epsilon_V}-\frac{29}{30}+
\mathcal{O}(\epsilon_V,\eta_\sigma,\eta_V)\right]\label{T00sfinal2}
\ee
The last term in eq. (\ref{T00sfinal2}) is completely determined
by the trace anomaly of a minimally coupled scalar field in de
Sitter space time\cite{BD,sanchez,duff,dowker}.

\subsection{Tensor perturbations}

Tensor perturbations correspond to massless fields with two physical
polarizations. The quantum fields are written as
$$
h^i_j(\vec x,\eta) = \frac{1}{C(\eta) \; M_{Pl} \sqrt{2 \; \Omega}}
\sum_{\lambda=\times,+} \sum_{\vec{k}}\,  e^{i \vec{k}\cdot \vec{x}}
 \; \epsilon^i_j(\lambda,\vec{k})
\left[a_{\lambda,\vec{k}} \; \,S_h(k,\eta)+
a^\dagger_{\lambda,\vec{k}} \; \,S^*_h(k,\eta) \right] \; ,
$$
\noindent where the operators
$ a_{\lambda,\vec{k}}, \; a^\dagger_{\lambda,\vec{k}} $ obey the usual
canonical commutation relations, and $\epsilon^i_j(\lambda,\vec{k})$
are the two independent traceless-transverse tensors constructed
from the two independent polarization vectors transverse to
$\hat{\bf{k}}$, chosen to be real and normalized such that
$\epsilon^i_j(\lambda,\vec{k})\, \; \epsilon^j_k(\lambda',\vec{k})=\delta^i_k
 \; \delta_{\lambda,\lambda'}$.

The mode functions $S_h(k,\eta)$ obey
the differential equation \be S^{''}_{h}(k,\eta)+\left[k^2-
\frac{\nu^2_{h}-\frac{1}{4}}{\eta^2} \right]S_{h}(k,\eta) = 0 ~~;~~
\nu_h = \frac{3}{2}+\epsilon_V +
\mathcal{O}(\epsilon^2_V,\eta^2_{\sigma},\eta^2_V,\epsilon_V
\eta_V)\label{Sten}
\ee
The solutions corresponding to the Bunch-Davies vacuum annihilated
by the operators $a_{\lambda,\vec{k}} $ are
\be \label{Stsol}S_{h}(k,\eta) = \frac{1}{2} \; \sqrt{-\pi\eta} \; \;
e^{i\frac{\pi}{2}(\nu_h+\frac{1}{2})} \; H^{(1)}_{\nu_h}(-k\eta) \; \, ,
\ee
The energy momentum tensor for gravitons only depends on
derivatives of the field $ h^i_j $ therefore its  expectation value in
the Bunch Davies (BD) vacuum does not feature infrared singularities in
the limit $\epsilon_V \rightarrow 0$. The absence of infrared
singularities in the limit of exact de Sitter space time, entails
that we can extract the leading contribution to the effective
potential from tensor perturbations by evaluating the expectation
value of $T_{00}$ in the BD vacuum in exact de Sitter
space time, namely by setting all slow roll parameters to zero. This
will yield the leading order in the slow roll expansion.

Because de Sitter space time is maximally symmetric, the expectation
value of the energy momentum tensor is given by\cite{weinberg,BD}
\be \label{trace}
\langle T_{\mu \nu} \rangle_{BD} = \frac{g_{\mu
\nu}}{4} \; \langle T^{ \alpha}_{\alpha } \rangle_{BD}
\ee
\noindent and $ T^{ \alpha}_{\alpha }$ is a space-time constant,
therefore the energy momentum tensor is manifestly covariantly
conserved. Of course, in a quantum field theory there emerge
ultraviolet divergences and the regularization procedure must be
compatible with the maximal symmetry. A large body of work has been
devoted to study the trace anomaly in de Sitter space time
implementing a variety of powerful covariant regularization methods that
preserve the symmetry\cite{duff,BD,dowker,hartle,fujikawa,sanchez}
yielding a renormalized value of the expectation value of the
$ \langle T_{\mu \nu} \rangle_{BD} $
given by eq. (\ref{trace}). Therefore, the full
energy momentum tensor is completely determined by the trace
anomaly \cite{BD,sanchez,duff}.

The contribution to the trace anomaly from gravitons has been
given in refs.\cite{duff,sanchez,BD}, it is \be \langle T^{
\alpha}_{\alpha } \rangle_t = -\frac{717}{80 \; \pi^2} \;  H^4_0
\label{traza} \ee From this result, we conclude that \be \langle
T^0_{0} \rangle_t = -\frac{717}{320 \;  \pi^2} H^4_0
\label{T00grav} \ee This result differs by a numerical factor from
that obtained in ref.\cite{finelli}, presumably the difference is
a result of a different regularization scheme.

\subsection{Fermion fields}

The Dirac equation in the FRW geometry is given by [see eq.(\ref{ecdi})],
\be
\Bigg[i \; {\not\!{\partial}}-M_\Psi[\Phi_0] \; C(\eta) \Bigg]
\left(C^{\frac{3}{2}}{\Psi({\vec x},\eta)}\right) = 0 \; .
\ee
The solution $ \Psi({\vec x},\eta) $ can be expanded in spinor mode
functions as
\be
\Psi(\vec{x},\eta) = \frac{1}{C^\frac{3}{2}(\eta) \;  \sqrt{\Omega}}
\sum_{\vec{k},\lambda}\,  e^{i \vec{k}\cdot
\vec{x}}\left[b_{\vec{k},\lambda}\, U_{\lambda}(\vec{k},\eta)+
d^{\dagger}_{-\vec{k},\lambda}\, V_{\lambda}(-\vec{k},\eta)\right] \; ,
\label{psiex}
\ee
where the spinor mode functions $U,V$ obey the  Dirac equations
\bea
\Bigg[i \; \gamma^0 \;  \partial_\eta - \vec{\gamma}\cdot \vec{k}
-M(\eta) \Bigg]U_\lambda(\vec{k},\eta) & = & 0 \label{Uspinor} \\
\Bigg[i \; \gamma^0 \;  \partial_\eta + \vec{\gamma}\cdot \vec{k} -M(\eta)
\Bigg]V_\lambda(\vec{k},\eta) & = & 0 \label{Vspinor}
\eea
\noindent and
\be
M(\eta) \equiv M_\Psi[\Phi_0]  \;  C(\eta)\label{Fmass}
\ee
Following the method of refs.\cite{boyarel,baacke}, it proves
convenient to write
\bea
U_\lambda(\vec{k},\eta) & = & \Bigg[i \; \gamma^0 \;  \partial_\eta -
\vec{\gamma}\cdot \vec{k} +M(\eta)
\Bigg]f_k(\eta)\, \mathcal{U}_\lambda \label{Us}\\
V_\lambda(\vec{k},\eta) & = & \Bigg[i \; \gamma^0 \;  \partial_\eta +
\vec{\gamma}\cdot \vec{k} +M( \eta)
\Bigg]g_k(\eta)\,\mathcal{V}_\lambda \label{Vs}
\eea
\noindent with $\mathcal{U}_\lambda;\mathcal{V}_\lambda$ being
constant spinors\cite{boyarel,baacke} obeying
\be
\gamma^0 \; \mathcal{U}_\lambda  =  \mathcal{U}_\lambda
\label{Up} \qquad , \qquad
\gamma^0 \;  \mathcal{V}_\lambda  =  -\mathcal{V}_\lambda
\ee
The mode functions $f_k(\eta);g_k(\eta)$ obey the following
equations of motion
\bea \left[\frac{d^2}{d\eta^2} +
k^2+M^2(\eta)-i \; M'(\eta)\right]f_k(\eta) & = & 0 \label{feq}\\
\left[\frac{d^2}{d\eta^2} + k^2+M^2(\eta)+i \; M'(\eta)\right]g_k(\eta)
& = & 0 \label{geq}
\eea
Neglecting the derivative of $\Phi_0$ with respect to time, namely
terms of order $\sqrt{\epsilon_V}$ and higher, the equations of
motion for the mode functions are given by
\bea \left[\frac{d^2}{d\eta^2} + k^2-\frac{\nu^2_+
-\frac{1}{4}}{\eta^2}\right]f_k(\eta) & = & 0 \label{fplu}\\
\left[\frac{d^2}{d\eta^2} + k^2-\frac{\nu^2_-
-\frac{1}{4}}{\eta^2}\right]g_k(\eta) & = & 0 \label{gmin}
\eea
\noindent where
$$
\nu_{\pm} = \frac{1}{2}\pm i  \;
\frac{M_\Psi[\Phi_0]}{H_0} 
$$
The scalar product of the spinors $U_\lambda(\vec
k,\eta), \; V_\lambda(\vec k,\eta)$ yields
\bea
U^\dagger_\lambda(\vec k,\eta) \; U_{\lambda'}(\vec k,\eta) & = &
\mathcal{C}^+(k) \; \delta_{\lambda,\lambda'} \cr \cr 
V^\dagger_\lambda(\vec k,\eta) \; V_{\lambda'}(\vec k,\eta) & = &
\mathcal{C}^-(k) \; \delta_{\lambda,\lambda'} \nonumber 
\eea
\noindent where
\bea \mathcal{C}^+(k) & = &
f^{*'}_k(\eta) \; f'_k(\eta)+\left(k^2+M^2(\eta)\right)
f^*_k(\eta) \; f_k(\eta)+i \; M(\eta)\left(f'_k(\eta) \; f^*_k(\eta)-
f_k(\eta) \; f^{*'}_k(\eta)\right)\cr \cr
\mathcal{C}^-(k) & = &
g^{*'}_k(\eta) \; g'_k(\eta)+\left(k^2+M^2(\eta)\right)
g^*_k(\eta) \; g_k(\eta)-i \; M(\eta)\left(g'_k(\eta) \; g^*_k(\eta)-
g_k(\eta) \; g^{*'}_k(\eta)\right) \nonumber 
\eea
\noindent are constants of motion by dint of the equations of motion
for the mode functions $f_k(\eta), \; g_k(\eta)$. The normalized spinor
solutions of the Dirac equation are therefore given by
\bea
&& U_\lambda(\vec k,\eta) =
\frac{1}{\sqrt{\mathcal{C}^+(k)}}\left[i \;
f'_k(\eta)-\vec{\gamma}\cdot\vec{k} \; f_k(\eta)+M(\eta) \; f_k(\eta)
\right] \,\mathcal{U}_\lambda  \cr \cr 
&& V_\lambda(\vec k,\eta) =
\frac{1}{\sqrt{\mathcal{C}^-(k)}}\left[-i \;
g'_k(\eta)+\vec{\gamma}\cdot\vec{k} \; g_k(\eta)+M(\eta) \; g_k(\eta)
\right] \,\mathcal{U}_\lambda \nonumber 
\eea
We choose the solutions of the mode equations
(\ref{fplu})-(\ref{gmin}) to be
\be
f_k(\eta) = \sqrt{\frac{-\pi k \eta}{2}}\;
e^{i\frac{\pi}{2}(\nu_+ + \frac{1}{2})}\;H^{(1)}_{\nu_+}(-k\eta)
\label{fksol} \qquad , \qquad
g_k(\eta) = \sqrt{\frac{-\pi k \eta}{2}}\;
e^{-i\frac{\pi}{2}(\nu_- + \frac{1}{2})}\;H^{(2)}_{\nu_-}(-k\eta)
\ee
We also choose the Bunch-Davies vacuum state such that $b_{\vec
k,\lambda}|0>_{BD}=0;d_{\vec k,\lambda}|0>_{BD}=0$. The choice of
the mode functions eq.(\ref{fksol}) yield the following
normalization factors
$$
\mathcal{C}^+(k) = \mathcal{C}^-(k) = 2 \; k^2 \; . 
$$
The energy momentum tensor for a spin $1/2$ field is given by\cite{BD}
$$ 
T_{\mu \nu} = \frac{i}{2}\left[\overline{\Psi} \gamma_{(\mu}
\stackrel{\leftrightarrow}{\mathcal{D}}_{\nu)}\Psi \right]\,
$$ 
\noindent and its expectation value in the Bunch-Davis vacuum
is equal to
$$
\langle T^0_{0} \rangle_{BD} = \frac{2}{C^4(\eta)}\int
\frac{d^3k}{(2\pi)^3}\Bigg\{M(\eta)-
\mathrm{Im}\left[g'_k(\eta)g^*_k(\eta)\right] \Bigg\}
$$
where $ M(\eta) $ and $ g_k(\eta) $ are given by eqs.(\ref{Fmass})
and (\ref{fksol}), respectively. It is clear that this energy
momentum tensor does not feature any infrared sensitivity because
the index of the Bessel functions is $\nu_\pm \approx 1/2$. Of
course this is expected since fermionic fields cannot feature
large amplitudes due to the Pauli principle.

A lengthy computation using covariant point splitting
regularization yields the following result \be \langle T^0_{0}
\rangle_{\Psi} = \frac{11 \, H^4_0}{960 \; \pi^2}\left\{1+
\frac{120}{11}\mathcal{M}^2\left(\mathcal{M}^2+1 \right) \left[ -
\mathrm{Re}\,\psi(2+i \mathcal{M}) - \frac{19}{12}-\gamma-2
\ln2\right] \right\} \quad , \quad
\mathcal{M}\equiv\frac{M_\Psi[\Phi_0]}{H_0} \label{T00fermi} \ee
The first term in the bracket in eq.(\ref{T00fermi}) is recognized
as the trace anomaly for fermions and is the only term that
survives in the massless
limit\cite{duff,hartle,BD,fujikawa,dowker,sanchez}. For light
fermion fields, $\mathcal{M} \ll 1$, and the leading contribution
to the energy momentum tensor is completely determined by the
trace anomaly, hence in this limit the contribution to the
covariantly regularized effective potential from (Dirac) fermions
is given by
$$ 
\langle T^0_{0} \rangle_{\Psi} = \frac{11 \, H^4_0}{960 \;
\pi^2}\left[ 1 + \mathcal{O}(\mathcal{M}^2)\right]
$$
This result is valid for Dirac fermions and it must be divided by a factor
2 for Weyl or Majorana fermions.

\subsection{Summary}

In summary, we find that the effective potential at one-loop is given by,
$$
\delta V(\Phi_0)  = \frac{H^4_0}{(4 \; \pi)^2} \Bigg[\frac{\eta_V-4 \;
\epsilon_V}{\eta_V-3 \; \epsilon_V}+ \frac{3\,\eta_\sigma}{\eta_\sigma-\epsilon_V}+
\mathcal{T}_{\Phi}+ \mathcal{T}_s+\mathcal{T}_t +\mathcal{T}_{\Psi}+
\mathcal{O}(\epsilon_V,\eta_V,\eta_\sigma,\mathcal{M}^2) \Bigg] \; ,
$$
\noindent where $(s,t,\sigma,\Psi)$ stand for the contributions
of the scalar metric, tensor fluctuations, light boson field $\sigma$ and light
fermion field $\Psi$, respectively. We have
\bea
\mathcal{T}_{\Phi} & = & \mathcal{T}_s = -\frac{29}{30} \cr \cr
\mathcal{T}_t & = & -\frac{717}{5} \label{Tt}\\
\mathcal{T}_{\Psi} & = & \frac{11}{60}\nonumber 
\eea
The terms that feature the \emph{ratios } of combinations of slow
roll parameters arise from the infrared or superhorizon
contribution from the scalar density perturbations and scalar
fields $\sigma$ respectively. The terms $\mathcal{T}_{s,t,\Psi}$
are completely determined by the trace anomalies of scalar,
graviton and fermion fields respectively. Writing $H^4_0 =
V(\Phi_0) \;  H^2_0/[3 \; M^2_{Pl}]$ we can finally write the effective
potential to leading order in slow roll
\be
V_{eff}(\Phi_0) = V(\Phi_0)\Bigg[1+ \frac{H^2_0}{3 \; (4 \; \pi)^2 \;
M^2_{Pl}}\Bigg( \frac{\eta_V-4\,\epsilon_V}{\eta_V-3 \; \epsilon_V}+
\frac{3\,\eta_\sigma}{ \eta_\sigma-\epsilon_V}-\frac{2903}{20}\Bigg)
\Bigg] \label{Veffin}
\ee
There are several remarkable aspects of this result:

i) the infrared
  enhancement as a result of the near scale invariance of scalar
  field fluctuations, both from scalar density perturbations as
  well as from a light scalar field, yield corrections of \emph{zeroth
  order in slow roll}. This is a consequence of the fact that
  during slow roll the particular combination $ \Delta_\sigma = \eta_\sigma-
\epsilon_V $ of slow roll parameters yield a natural infrared cutoff.

ii) the final one   loop contribution to the effective potential displays the
  effective field theory dimensionless parameter $H^2_0/M^2_{Pl}$
confirming our previous studies\cite{nuestros,nuestronor},

iii) the last term is completely
  determined by the trace anomaly, a purely geometric result of the
short distance properties of the theory.

\section{Quantum Corrections to the Curvature and Tensor Fluctuations}

The quantum corrections to the effective potential lead to quantum corrections
to the amplitude of scalar and tensor fluctuations.

The scalar curvature and tensor fluctuations in the slow-roll regime are
giving by the formulas\cite{liddle}
\be\label{amplis}
|{\Delta}_{k}^{(S)}|^2  =
\frac{1}{8 \, \pi^2 \; \epsilon_V} \; \left( \frac{H}{M_{Pl}}\right)^2
\quad ,  \quad |{\Delta}_{k}^{(T)}|^2  =
\frac{1}{2 \, \pi^2} \; \left( \frac{H}{M_{Pl}}\right)^2 \; .
\ee
where $ H $ stands for the Hubble parameter and $\epsilon_V$ is given by
eq.(\ref{etav}).

We can include the leading quantum corrections in eq.(\ref{amplis}) replacing
in it $ H $ and $ \epsilon_V $ by the corrected parameters $ H_{eff} $ and
$ \epsilon_{eff} $. That is,
\be\label{hyep}
H_{eff}^2 = H_0^2 + \delta H^2 \quad ,  \quad
\epsilon_{eff} = \epsilon_V +  \delta \epsilon_V
\ee
with
\be\label{efe}
H_{eff}^2 = \frac{V_{eff}(\Phi_0)}{3 \;  M_{Pl}^2} \quad , \quad
\epsilon_{eff} =\frac{M^2_{Pl}}{2} \;
\left[\frac{V_{eff}^{'}(\Phi_0)}{V_{eff}(\Phi_0)} \right]^2  \; ,
\ee
and where $ V_{eff}(\Phi_0) $ is given by eq.(\ref{Veffin}).
We thus obtain,
\bea\label{dhe}
&&\frac{\delta H^2}{H_0^2} = \frac13\left(\frac{H_0}{4 \; \pi \; M_{Pl}}\right)^2
\Bigg[ \frac{\eta_V-4\,\epsilon_V}{\eta_V-3 \, \epsilon_V}+
\frac{3\,\eta_\sigma}{\eta_\sigma-\epsilon_V}-\frac{2903}{20}\Bigg] \; ,
\\ \cr
&&\frac{\delta\epsilon_V}{\epsilon_V} =  \frac23\left(\frac{H_0}{4 \; \pi
\; M_{Pl}}\right)^2 \left\{ \frac{\xi_V + 12 \; \epsilon_V \left( 2 \;
\epsilon_V - \eta_V \right)}{2 \; \left( \eta_V - 3 \; \epsilon_V\right)^2}
+ \frac{3 \; \eta_\sigma}{\left(\eta_{\sigma} -
\epsilon_V\right)^2} \left[\eta_\sigma + \eta_V - 2 \; \epsilon_V
 - \sqrt{2 \; \epsilon_V} \; M_{Pl} \;
\frac{d \log M_{\sigma}[\Phi_0]}{d \Phi_0}  \right] -\frac{2903}{20}
\right\} \nonumber
\eea

Inserting eq.(\ref{dhe}) into eqs.(\ref{hyep}) and (\ref{efe}) yields after
calculation, for the scalar perturbations,
\bea\label{delref}
&& |{\Delta}_{k,eff}^{(S)}|^2  =
|{\Delta}_{k}^{(S)}|^2  \left[ 1 -
\frac{\delta\epsilon_V}{\epsilon_V} + \frac{\delta H^2}{H^2} \right]=\cr \cr
&& = |{\Delta}_{k}^{(S)}|^2  \left\{ 1
- \frac13 \left(\frac{H_0}{4 \; \pi \; M_{Pl}}\right)^2
\left[\frac{\xi_V + 12 \; \epsilon_V^2 - \eta_V^2 - 5 \;  \epsilon_V \;
\eta_V}{\left(\eta_V - 3 \; \epsilon_V\right)^2} +
\right. \right.  \cr \cr
&&\left.  \left. +\frac{3 \; \eta_\sigma}{\left(\eta_{\sigma} -
\epsilon_V\right)^2} \left[\eta_\sigma - 3 \; \epsilon_V +
2 \; \eta_V - 2\;  \sqrt{2 \; \epsilon_V} \;
M_{Pl} \; \frac{d \log M_{\sigma}[\Phi_0]}{d \Phi_0}  \right]
- \frac{2903}{20}\right] \right\} \; ,
\eea
and for the tensor perturbations,
\bea\label{delref2}
&& |{\Delta}_{k,eff}^{(T)}|^2  =|{\Delta}_{k}^{(T)}|^2  \left[ 1
+ \frac{\delta H^2}{H^2} \right]=\cr \cr
&&=|{\Delta}_{k}^{(T)}|^2  \left\{ 1
+ \frac13 \left(\frac{H_0}{4 \; \pi \; M_{Pl}}\right)^2
\Bigg[ \frac{\eta_V-4\,\epsilon_V}{\eta_V-3 \, \epsilon_V}+
\frac{3\,\eta_\sigma}{\eta_\sigma-\epsilon_V}-\frac{2903}{20}\Bigg]
 \right\} \; .
\eea
where $ M_{\sigma}[\Phi_0] $ and $ \eta_\sigma $ are given by
eqs.(\ref{sigmamass}) and (\ref{etasig}), respectively.

\bigskip

The case when  the field $\sigma$ is  much lighter than the
inflaton permit simplifications, since
$$
\eta_\sigma \sim \left(\frac{m_\sigma}{m_{inflaton}} \right)^2 \;\eta_V \; ,
$$
for $ m_\sigma^2 \ll m_{inflaton}^2 $, we can neglect terms
proportional to  $ \eta_\sigma $ in the expressions for  $
|{\Delta}_{k}^{(S)}|^2 $ and $  |{\Delta}_{k,eff}^{(T)}|^2 $. In
this case the quantum corrections to the power spectra obtain a
 particularly illuminating expression when the slow-roll parameters
in eqs.(\ref{delref})-(\ref{delref2}) are written in terms of the
CMB observables $ n_s, \; r $ and the spectral running of the
scalar index using 
\bea\label{gorda} 
&&\epsilon_V  = \frac{r}{16}
\quad , \quad \eta_V   =\frac12\left( n_s - 1 + \frac{3}{8} \, r
\right) \quad , \cr \cr && \xi_V = \frac{r}4 \left(n_s - 1 +
\frac{3}{16} \, r \right) - \frac12\frac{dn_s}{d \ln k} \quad ,
\quad \eta_V - 3 \; \epsilon_V = \frac12 (n_s - 1) \; . \eea We
find from eqs.(\ref{delref})-(\ref{delref2}), \bea &&
|{\Delta}_{k,eff}^{(S)}|^2  =  |{\Delta}_{k}^{(S)}|^2  \left\{ 1+
\frac23 \left(\frac{H_0}{4 \; \pi \; M_{Pl}}\right)^2
\left[1+\frac{\frac38 \; r \; (n_s - 1) + 2 \; \frac{dn_s}{d \ln
k}}{(n_s - 1)^2} + \frac{2903}{40} \right] \right\} \cr \cr &&
|{\Delta}_{k,eff}^{(T)}|^2  =|{\Delta}_{k}^{(T)}|^2  \left\{ 1
-\frac13 \left(\frac{H_0}{4 \; \pi \; M_{Pl}}\right)^2
\left[-1+\frac18 \; \frac{r}{n_s - 1}+ \frac{2903}{20}
\right]\right\} \; . \eea We see that the anomalies contribution $
\frac{2903}{40} = 72.575 $ and  $ \frac{2903}{20} = 145.15 $
presumably dominate both quantum corrections. The other terms are
generally expected to be smaller than these large contributions
from the anomalies. These anomalous contributions are dominated in
turn by the tensor part [see eq.(\ref{Tt})]. Only fermions give
contributions with the opposite sign. However, one needs at least
$783$ species of (Dirac) Fermions to compensate for the tensor
part.

These quantum corrections also affect the ratio $ r $ of tensor/scalar
fluctuations as follows,
\be
r_{eff} \equiv \frac{|{\Delta}_{k,eff}^{(T)}|^2}{|{\Delta}_{k,eff}^{(S)}|^2}
= r \;  \left\{ 1-\frac13 \left(\frac{H_0}{4 \; \pi \; M_{Pl}}\right)^2
\left[1+\frac{\frac38 \; r \; (n_s - 1) +
\frac{dn_s}{d \ln k}}{(n_s - 1)^2} + \frac{8709}{20} \right] \right\}
\ee
We expect this quantum correction to the ratio to be negative
as the anomaly contribution  dominates: $ \frac{8709}{20} = 435.45 $.

\bigskip

Therefore, the quantum corrections {\bf enhance} the scalar curvature
fluctuations while they {\bf reduce} the tensor fluctuations
as well as their ratio $r$. The quantum corrections are small, of the order
$ \left(\frac{H_0}{M_{Pl}}\right)^2 $, but it is interesting to see that
the quantum effects are dominated by the trace anomalies
and they correct both fluctuations in a definite direction.

\section{Conclusions}

Motivated by the premise that forthcoming CMB observations may
probe the inflationary potential, we study its quantum corrections
from scalar and tensor metric perturbations as well as those from
one light scalar and one light (Dirac) fermion field generically
coupled to the inflaton. The reason for this study is that the measurements
probe the {\bf full} effective inflaton potential, namely
the classical potential plus its quantum corrections.
We have focused on obtaining the quantum
corrections to the effective potential during the cosmologically
relevant quasi-de Sitter stage of slow roll inflation. Both, scalar
metric fluctuations, as well as those from a light scalar field,
feature infrared enhancements as a consequence of the nearly scale
invariance of their power spectra. A combination of slow roll
parameters appropriate for each case provides a natural infrared
regularization. We find that to leading order in slow roll, there
is a {\it clean and unambiguous} separation between the contributions to
the effective potential from superhorizon modes of the scalar
metric perturbations as well as the scalar field, and those
from subhorizon modes. Only the contributions to the total energy
momentum tensor from curvature perturbations and the light scalar
field feature an infrared enhancement, while those from
gravitational waves and fermions {\bf do not} feature any infrared
sensitivity.

In all cases, scalar metric, tensor, light scalar and fermion fields,
the contribution from subhorizon modes is determined by the trace anomaly,
while the contribution from superhorizon
modes, only relevant for curvature and scalar field perturbations,
are infrared enhanced as the inverse of a combination of slow roll parameters 
which measure the departure from scale invariance in each case.

The one loop effective potential to leading order in slow roll is
given by eq. (\ref{Veffin}). The last term, independent of the
slow roll parameters, is completely determined by the trace
anomalies of scalar, tensor and fermionic fields
(therefore solely determined by the space-time geometry), while the first
term reveals the hallmark infrared enhancement of superhorizon
fluctuations. Using this result we have obtained the one-loop
quantum corrections to the amplitude of (scalar) curvature and tensor
perturbations in terms of the CMB observables $ n_s, \; r $ and 
$ d n_s/d \ln k$ and the total trace anomaly $ \mathcal{T} $
of the different fields.

As we anticipated in ref.\cite{nuestros,nuestronor}, the strength of
the one loop corrections is determined by the effective field
theory parameter $(H_0/M_{Pl})^2 $. While this quantity is
observationally of $\mathcal{O}(10^{-10})$, there is an important
message in this result: the {\bf robustness} of slow-roll inflation as
well as the reliability of the effective field theory description.

There is a simple interpretation of the above result: in the
effective field theory approach, the `classical' inflaton
potential $V(\Phi_0)$ includes contributions from integrating out
the fields with scales much heavier than the scale of inflation
$H_0$.

The contribution to the energy momentum from light fields, yield
the effective potential, however for fields with mass scales $\ll
H_0$ the dominant scale in the problem is $H_0$ and on dimensional
grounds the contribution to the covariantly renormalized energy
momentum tensor must be $\propto H^4_0$. This argument would fail
in the presence of \emph{infrared} divergences, and indeed the
mass terms from curvature perturbations and from the scalar
field $\sigma$ feature an infrared enhancement because of their
nearly scale invariant power spectrum. The mass term is of first
order in the slow roll, however the infrared enhancement brings
about a denominator which is also of first order in slow roll
yielding a ratio  which is of zeroth order in slow roll. Hence, this
remarkable result validates the simple power counting that
yields the overall scale $H^4_0$ for the one loop correction. 

An important bonus of the slow roll approximation is that the
contributions from superhorizon and subhorizon modes can be
\emph{unambiguously separated} and the latter are completely determined
by the trace anomaly, a purely geometrical result which only
depends on the short distance (ultraviolet) properties.
Quantum trace anomalies of the energy momentun tensor in
gravitational fields constitute a nice and important chapter of
QFT in curved backgrounds, (see \cite{BD} and references therein).
Our results here show that these trace anomalies dominate the
quantum corrections to a relevant cosmological problem:
the primordial power spectrum of curvature and tensor fluctuations.

\begin{acknowledgments} D.B.\ thanks the US NSF for support under
grant PHY-0242134,  and the Observatoire de Paris and LERMA for
hospitality during this work. He also thanks L. R. Abramo for
interesting discussions. This work is supported in part by the
Conseil Scientifique de l'Observatoire de Paris through an `Action
Initiative, BQR'.
\end{acknowledgments}

\end{document}